\documentclass[english]{IEEEtran}
\usepackage{amsmath,balance}
\usepackage[T1]{fontenc}
\usepackage[latin9]{inputenc}
\usepackage{amsthm}
\usepackage{amsmath}
\usepackage{amssymb}
\usepackage{graphicx}
\usepackage{balance}
\usepackage{epstopdf}
\usepackage{algpseudocode}
\usepackage{algorithm}
\usepackage{bbold}

\makeatletter

\theoremstyle{plain}
\newtheorem{thm}{\protect\theoremname}
\theoremstyle{remark}
\newtheorem{rem}[thm]{\protect\remarkname}

\makeatother

\usepackage{babel}
\providecommand{\remarkname}{Remark}
\providecommand{\theoremname}{Theorem}
\graphicspath{{Figures/}}

\begin{document}

\title{A Quaternion Frequency and Phasor Estimator for Three-Phase Power Distribution Networks}

\author{Sayed Pouria Talebi and Danilo P. Mandic\\
\thanks{Sayed Pouria Talebi and Danilo P. Mandic are with the department of Electrical and Electronic Engineering, Imperial College London, SW7 2AZ, U.K. E-mail:\{s.talebi12, d.mandic\}@imperial.ac.uk.}}
\maketitle

\maketitle
\begin{abstract}
The multi-dimensional nature of quaternions allows for the full characterization of three-phase power systems. This is achieved through the use of quaternions to provide a unified framework for incorporating voltage measurements from all the phases of a three-phase system and then employing the recently introduced $\mathbb{HR}$-calculus to derive a state space estimator based on the quaternion extended Kalman filter (QEKF). The components of the state space vector are designed such that they can be deployed for adaptive estimation of the system phasors. Finally, the proposed algorithm is validated through simulations using both synthetic and real-world data, which indicate that the developed quaternion frequency estimator can outperform its complex-valued counterparts.
\end{abstract}

\section{Introduction}

The power grid is designed to operate optimally at a nominal frequency and in a balanced fashion~\cite{Baggini}. Deviations from the nominal nominal frequency and unbalanced operating conditions can adversely affect the performance of different components of the power grid such as compensators and loads~\cite{Modelling-effects}-\cite{Assessment-voltage}, resulting in harmful operating conditions that can propagate throughout the network. Therefore, monitoring frequency and phasor information of three-phase power distribution networks in real-time is a prerequisite for ensuring nominal operating conditions. 

Moreover, the system frequency reveals essential information about the dynamics of the power grid. For example, a rising system frequency indicates that power generation has surpassed power consumption and a falling system frequency is indicative of power consumption exceeding power generation. Thus, reliable frequency estimators are an essential part of power grid management systems that maintain the balance between power generation and consumption~\cite{Proc-Mag}.

The importance of frequency estimation in three-phase power systems has prompted the introduction of a number of algorithms for this purpose including phase-locked loops (PLL)~\cite{PLL-I}-\cite{PLL}, a recursive Newton-type frequency and spectrum estimation algorithm~\cite{IRNT}-\cite{VPLF}, approaches based on the least squares and least mean square algorithms~\cite{key-ADD}-\cite{PSF}, and a Fourier transform-based method for estimating the main frequency power systems~\cite{RT-FFT}. Among these algorithms, approaches based on the Kalman and extended Kalman filters have been shown to be advantageous~\cite{FED}-\cite{KFE}. Most approaches for estimating the frequency of three-phase power systems are either based on real-valued algorithms and use measurements from a single phase or use the Clarke transform to map voltage measurements from all three phases on top the complex domain and use standard linear complex-valued algorithms and are shown to be accurate under nominal operating conditions; however, under unbalanced operating conditions these approaches can lead to non-unique solutions and suffer from large oscillatory errors~\cite{DyM}-\cite{Mojiri}. To address this issue, frequency estimators based on widely linear complex signal processing algorithms have been developed~\cite{Y}-\cite{D}.

Although mapping the three-phase voltages onto the complex domain allows for the use of well established complex-valued signal processing techniques; however, complex numbers lack the dimensionality necessary to model three-phase systems which leads to partial loss of information specially under unbalanced operating conditions. This loss of information not only effects the performance of complex-valued algorithms, but also makes it impossible to retrieve the phasor information of the three-phase system. Quaternions provide a natural presentation for three and four-dimensional signals and have gained popularity in a variety of engineering application~\cite{FAT}-\cite{book}. Moreover, the recent introduction of the $\mathbb{HR}$-calculus~\cite{Gradient} and augmented quaternion statistics~\cite{Aug-Quat}-\cite{Prop-Wide} have led a resurgence in quaternion-valued signal processing and have inspired the introduction of a number of quaternion-valued signal processing algorithms including a class of quaternion Kalman filters~\cite{QKF}.

In this work, we re-drive the quaternion frequency estimator presented in our previous work~\cite{Me}. The new derivation provides a better insight to the physical interpretation of the elements of the state space vector that is then exploited to estimate the phasors of each phase. The resulting estimator can fully characterize three-phase systems, operates optimally under both balanced and unbalanced conditions, and eliminates the need to use the Clark transform. The developed algorithm is validated through simulation using both synthetic data and real-world data recordings where it is shown that it can outperform its complex-valued counterparts. 

\section{Background}

\subsection{Three-phase power systems}

The instantaneous voltages of each phase in a three-phase power system is given by~\cite{Clarke}
\begin{equation}
\begin{aligned}
v_{a,n}=&V_{a,n}\text{sin}\left(2\pi f \Delta T n + \phi_{a,n}\right)
\\
v_{b,n}=&V_{b,n}\text{sin}\Big(2\pi f \Delta T n +  \phi_{b,n} +  \frac{2 \pi}{3}\Big)
\\
v_{c,n}=&V_{c,n}\text{sin}\Big(2\pi f \Delta T n +  \phi_{c,n} +  \frac{4 \pi}{3}\Big)
\end{aligned}
\label{eq:thee-phase voltages}
\end{equation}
where $V_{a,n}$, $V_{b,n}$, and $V_{c,n}$ are the instantaneous amplitudes, $\phi_{a,n}$, $\phi_{b,n}$, and $\phi_{c,n}$ represent the instantaneous phases shifts, and $f$ is the system frequency, while $\Delta T = 1/f_{s}$ is the sampling interval with $f_{s}$ representing the sampling frequency. The Clarke transform, given by~\cite{Clarke} 
\[
\left[\begin{array}{c} v_{0,n}
\\
v_{\alpha,n}
\\
v_{\beta,n}
\end{array}\right]
=
\sqrt{\frac{2}{3}}
\left[\begin{array}{ccc}
\frac{\sqrt{2}}{2} & \frac{\sqrt{2}}{2} & \frac{\sqrt{2}}{2}
\\
1 & -\frac{1}{2} & -\frac{1}{2}
\\
0 & \frac{\sqrt{3}}{2} & -\frac{\sqrt{3}}{2}
\end{array}\right]\left[\begin{array}{c}
v_{a,n}
\\
v_{b,n}
\\
v_{c,n}
\end{array}\right]
\]
maps the three-phase system onto a new domain where it is conveniently represented by the complex-valued signal $v_{n}=v_{\alpha,n} + iv_{\beta,n}$, while $v_{0,n}$ is ignored in practical applications.

For a balanced three-phase system  $V_{n}=V_{a,n}=V_{b,n}=V_{c,n}$ and $\phi_{n}=\phi_{a,n}=\phi_{b,n}=\phi_{c,n}$, resulting in $v_{0,n}=0$; therefore, under balanced operating conditions 
\[
v_{n}=\sqrt{\frac{3}{2}}V_{n}e^{i(2\pi f \Delta T n + \phi_{n})}
\]
which can be expressed by the first order strictly linear regression 
\[
v_{n}=e^{i2\pi f \Delta T}v_{n-1}
\]
where the term $e^{i2 \pi f \Delta T}$ is referred to as the phase incrementing element. Moreover, the frequency of the system can then be estimated using standard linear complex extended Kalman filters employing the state space model given in Algorithm~\ref{Al:L-SS}, where $\varphi_{n}$ is the phase increment, $\boldsymbol{\varepsilon}_{n}$ the state evolution noise, and $\omega_{n}$ the observation noise~\cite{FED}.

\begin{algorithm}[!h]
\caption{Linear state space model (L-SS)}
State evolution equation: $\begin{bmatrix}\varphi_{n}\\ v_{n}\end{bmatrix}=\begin{bmatrix}\varphi_{n-1}\\ \varphi_{n-1} v_{n-1}\end{bmatrix}+\boldsymbol{\varepsilon}_{n}$ \vspace{0.5em}
\\
Observation equation: $v_{n}=\begin{bmatrix}0&1\end{bmatrix}\begin{bmatrix}\varphi_{n}\\ v_{n}\end{bmatrix}+\omega_{n}$ \vspace{0.5em}
\\  \vspace{5pt}
Estimate of frequency: $\hat{f}_{n}=\frac{1}{2 \pi \Delta T} \Im \left( \text{ln}\left(\varphi_{n} \right) \right)$\\ 
\vspace{-8pt}
\label{Al:L-SS}
\end{algorithm}

In practice a wide range of phenomena such as voltage sags, load imbalance, and faults in the transmission line can lead to the three-phase system operating in an unbalanced fashion~\cite{MHJBollen}. For a three-phase system operating under unbalanced conditions~\cite{Y}
\[
v_{n}=A_{n}e^{j(2 \pi f \Delta T n +\phi_{n})} + B_{n}e^{-j(2 \pi f \Delta T n + \phi_{n})}
\]
where 
\[
\begin{aligned}
A_{n}=&\frac{\sqrt{6}\left(  V_{a,n} + V_{b,n} + V_{c,n} \right)}{6}
\\
B_{n}=&\frac{\sqrt{6}\left(  2V_{a,n} - V_{b,n} - V_{c,n} \right)}{12} - j\frac{\sqrt{2}\left(   V_{b,n} - V_{c,n} \right)}{4} \cdot
\end{aligned}
\]
and all phase shifts are considered to be equal to $\phi_{n}$. Notice that $v_{n}$ comprises both a positive and a negative sequenced element. The presence of the negative sequenced element compromises the performance of strictly linear complex estimators, such as Algorithm~\ref{Al:L-SS}, and will lead to large oscillatory errors at twice the system frequency~\cite{Proc-Mag}.

In order to accommodate both balanced and unbalanced systems, $v_{n}$ can be expressed by employing the first order widely linear regression
\[
v_{n}=h_{n-1}v_{n-1}+g_{n-1}v^{*}_{n-1}
\]
where $h_{n}$ and $g_{n}$ are respectively the linear and conjugate weights~\cite{Y}. The fundamental frequency of both balanced and unbalanced three-phase systems can now be estimated by using a widely linear extended complex Kalman filter employing the state space model given in Algorithm~\ref{Al:WL-SS}, where $\boldsymbol{\varepsilon}^a_{n}$ and $\boldsymbol{\omega}^{a}_{n}$ are respectively the augmented state evolution and observation noise vectors~\cite{D}.

\begin{algorithm}[!h]
\caption{Widely linear state space model (WL-SS)}
State evolution equation: \[\begin{bmatrix}h_{n}\\g_{n}\\ v_{n} \\h^{*}_{n}\\g^{*}_{n}\\ v^{*}_{n}\end{bmatrix}=\begin{bmatrix}h_{n-1}\\g_{n-1}\\h_{n-1} v_{n-1}+g_{n-1} v^{*}_{n-1}\\h^{*}_{n-1}\\g^{*}_{n-1}\\h^{*}_{n-1} v^{*}_{n-1}+g^{*}_{n-1} v_{n}\end{bmatrix}+\boldsymbol{\varepsilon}^a_{n}\] \\ \vspace{5pt}
Observation equation: \[ \begin{bmatrix}v_{n}\\v^{*}_{n}\end{bmatrix}=\begin{bmatrix}
0 & 0 & 1 & 0 & 0 & 0\\ 
0 & 0 & 0 & 0 & 0 & 1
\end{bmatrix}\begin{bmatrix}h_{n}\\g_{n}\\ v_{n} \\h^{*}_{n}\\g^{*}_{n}\\ v^{*}_{n}\end{bmatrix}+\boldsymbol{\omega}^{a}_{n} \] \\  \vspace{5pt}
Estimate of frequency: 
$\hat{f}_{n}=\frac{1}{2 \pi \Delta T}\text{arcsin}\left(\Im\left(h_{n} + a_{n}\right)\right)$\\  \vspace{5pt}
where \[a_{n}=-j\Im\left(h_{n}\right)+j\sqrt{\Im^2\left(h_{n}\right)-\left|g_{n}\right|^2}\]\vspace{-4pt}
\label{Al:WL-SS}
\end{algorithm}

\subsection{Quaternions}

The skew-field of quaternions is a four-dimensional division algebra denoted by $\mathbb{H}$. A quaternion variable $q \in \mathbb{H}$ consists of a real part, $\Re(q)$,  and a three-dimensional imaginary part or pure quaternion, $\Im(q)$, which comprises three components $\Im_{i}(q)$, $\Im_{j}(q)$, and $\Im_{k}(q)$; hence, $q$ can be expressed as
\[
\begin{aligned}
q =\Re(q) + \Im(q) =&\Re(q) + \Im_{i}(q) + \Im_{j}(q) + \Im_{k}(g)
\\
=&q_{r} + iq_{i} + jq_{j} + kq_{k}
\end{aligned}
\]
where $q_{r},q_{i},q_{j},q_{k} \in \mathbb{R}$. The unit vectors $i$, $j$, and $k$ are the orthonormal basis for the quaternion imaginary subspace and obey the following product rules
\[
\begin{array}{c}
ij=k, jk=i, ki=j,
\\
 i^{2}=j^{2}=k^{2} = ijk = -1.
\end{array}
\]
The product of $q_{1}, q_{2}\in\mathbb{H}$ is given by
\begin{equation}
\begin{alignedat}{1}
q_{1}q_{2}= & \Re(q_{1})\Re(q_{2})+\Re(q_{1})\Im(q_{2})+\Re(q_{2})\Im(q_{1})
\\
&\hspace{1em}\hspace{1em}\hspace{1em}+\Im(q_{1})\times\Im(q_{2})-\big\langle\Im(q_{1}),\Im(q_{2})\big\rangle
\end{alignedat}
\label{eq:multiplication}
\end{equation}
where the symbols `$\langle\cdot,\cdot\rangle$' and `$\times$' denote the inner- and cross-products and the non-commutativity of the product is inherited from the vector product in (\ref{eq:multiplication}).

The quaternion conjugate of $q \in \mathbb{H}$ is defined as $q^{*}=\Re(q)-\Im(q)$, while the norm of $q$ is given by  
\[
|q| = \sqrt{qq^{*}} = \sqrt{q^{2}_{r}+q^{2}_{i}+q^{2}_{j}+q^{2}_{k}}.
\]
The quaternion conjugate is a special case of quaternion involutions. The involution of $q \in \mathbb{H}$ around $\eta \in \mathbb{H}$ is defined as~\cite{Q-invo}
\[
q^{\eta} \triangleq \eta q \eta ^{-1}
\]
which rotates the imaginary part of $q$ around $\Im(\eta)$ by $2\text{\hspace{1pt}arctan}\big(|\Im(\eta)|/\Re(\eta)\big)$ and has made quaternions ideal for modeling three-dimensional rotations (see~\cite{FAT}-\cite{book}, ~\cite{Q-invo} and references therein).

Quaternion involutions are also employed to express the real-valued components of the quaternion variable $q \in \mathbb{H}$, as~\cite{Gradient}-\cite{Me}
\begin{equation}
\begin{aligned}
q_{r} =& \frac{1}{4}\left(q + q^{i} + q^{j} + q^{k} \right) & q_{i} =&\frac{1}{4i}\left(q + q^{i} - q^{j} - q^{k}\right)
\end{aligned}
\label{eq:real-valued components}
\end{equation}
\[
\begin{aligned}
q_{j} =& \frac{1}{4j}\left(q - q^{i}+ q^{j} - q^{k} \right) & q_{k} =&\frac{1}{4k}\left(q - q^{i} - q^{j} + q^{k}\right)
\end{aligned}
\]
furthermore, the quaternion conjugate of $q$ can also be expressed by quaternion involutions as 
\begin{equation}
q^{*}=\frac{1}{2}\left(q^{i} + q^{j} + q^{k} - q \right).
\label{eq:quaternion conjugate}
\end{equation}

A quaternion $q \in \mathbb{H}$ can alternatively be expressed by its polar presentation given by~\cite{QFFT} 
\[
q = |q|e^{\xi \theta}=|q|\big(\text{cos}(\theta) + \xi \text{sin}(\theta) \big)
\]
where
\[
\xi = \frac{\Im(q)}{|\Im(q)|}, \hspace{1em} \theta = \text{atan}\left(\frac{|\Im(q)|}{\Re(q)}\right).
\]
Moreover, it is straightforward to prove that the $\text{sin}(\cdot)$ and $\text{cos}(\cdot)$ functions can be expressed as
\begin{equation}
\text{sin}(\theta)=\frac{1}{2\xi}\left(e^{\xi \theta} - e^{- \xi \theta}\right), \hspace{0.5em}\text{cos}(\theta)=\frac{1}{2}\left(e^{\xi \theta} + e^{- \xi \theta}\right)
\label{eq:quaternion sin and cos}
\end{equation} 
where $\xi^{2} = -1$\footnote{Note that in order to express the $\text{sin}(\cdot)$ and $\text{cos}(\cdot)$ functions in their polar from, as in (\ref{eq:quaternion sin and cos}), $\xi$ can be replace with an arbitrary normalized pure quaternion number~\cite{QFFT}.}.

Due to the restrictiveness of differentiability conditions of quaternion-valued functions (Cauchy-Riemann-Fueter condition), given by~\cite{Q-analysis}-\cite{On-HR}
\[
\frac{\partial f(\mathbf{q})}{\partial \mathbf{q}^{*}} = \frac{1}{4} \Big(\frac{\partial f}{\partial \mathbf{q}_{r}} + i\frac{\partial f}{\partial \mathbf{q}_{i}} + j\frac{\partial f}{\partial \mathbf{q}_{j}} + k\frac{\partial f}{\partial \mathbf{q}_{k}}  \Big)=0
\]
the calculation of quaternion derivatives has been a major stumbling block in the derivation of signal processing algorithms in the quaternion domain. One elegant solution to this problem is presented in the form of the $\mathbb{HR}$-calculus~\cite{Gradient},\cite{On-HR}, where based on the expressions in (\ref{eq:real-valued components}) a quaternion function $f\big(\mathbf{q}_{r},\mathbf{q}_{i},\mathbf{q}_{j},\mathbf{q}_{k}\big):~\mathbb{H}^{N}~\rightarrow~\mathbb{H}$ is expressed as a function of the augmented quaternion variable, $\mathbf{q}^{a}=[\mathbf{q}, \mathbf{q}^{i}, \mathbf{q}^{j},\mathbf{q}^{k}]^{T}$ as $f(\mathbf{q}^{a}):\mathbb{H}^{4N}\rightarrow\mathbb{H}$~\cite{Gradient},\cite{On-HR}, then the $\mathbb{HR}$-calculus exploits the duality between $[\mathbf{q}_{r},\mathbf{q}_{i},\mathbf{q}_{j}, \mathbf{q}_{k}]^{T}~\in~\mathbb{R}^{4N}$ and  $[\mathbf{q},\mathbf{q}^{i},\mathbf{q}^{j},\mathbf{q}^{k}]^{T} \in \mathbb{H}^{4N}$ to establish a framework for calculating the derivatives of $f$ directly in the quaternion domain. The isomorphism between $[\mathbf{q}_{r},\mathbf{q}_{i},\mathbf{q}_{j}, \mathbf{q}_{k}]^{T} \in \mathbb{R}^{4N}$ and  $[\mathbf{q},\mathbf{q}^{i},\mathbf{q}^{j},\mathbf{q}^{k}]^{T} \in \mathbb{H}^{4N}$ has also been instrumental in the development of the augmented quaternion statistics that allow for a full second-order description of  quaternion random variables~\cite{Aug-Quat}-\cite{Prop-Wide}. 

The augmented quaternion statistics in conjunction with the $\mathbb{HR}$-calculus have led to the development of a class of quaternion Kalman filters, including the strictly linear quaternion extended Kalman filter (QEKF)~\cite{QKF}~that operates akin to its complex-valued counterpart, with the difference that the Jacobian of the state evolution and observation functions have to be calculated by the $\mathbb{HR}$-calculus. For example, $\frac{\partial q^{*}}{ \partial q}=-0.5$, which is a consequence of (\ref{eq:quaternion conjugate}) and is in contrast to the results in the complex domain. The state evolution and observation equations for the QEKF are given by 
\[
\begin{aligned}
\mathbf{x}_{n}=& f(\mathbf{x}_{n-1})+\mathbf{r}_{n}
\\
\mathbf{y}_{n}=& g(\mathbf{x}_{n})+ \mathbf{s}_{n}
\end{aligned}
\]
and the operations of the quaternion extended Kalman filter are summarized in Algorithm~\ref{Al:QEKF}, where $f(\cdot)$ and $g(\cdot)$ are quaternion-valued functions with Jacobian matrices $\mathbf{A}_{n}$ and $\mathbf{H}_{n}$ at time instant $n$, whereas $\mathbf{s}_{n}$ and $\mathbf{r}_{n}$ denote the state evolution and observation noise vectors with covariance matrices $\mathbf{C}_{\mathbf{s}_{n}}$ and $\mathbf{C}_{\mathbf{r}_{n}}$, while $\hat{\mathbf{x}}_{n|n-1}$ and $\hat{\mathbf{x}}_{n|n}$ represent the \textit{a priori} and \textit{a posteriori} estimates of $\mathbf{x}_{n}$.

\begin{algorithm}[!h]
\caption{Quaternion extended Kalman filter (QEKF)}
Initialize: $\hat{\mathbf{x}}_{0|0}$ and $\hat{\mathbf{M}}_{0|0}$ 

For $n=1,2,...$: \\ \vspace{5pt}
$\hat{\mathbf{x}}_{n|n-1}=f(\hat{\mathbf{x}}_{n-1|n-1})$ \\  \vspace{5pt}
$\hat{\mathbf{M}}_{n|n-1}=\mathbf{A}_{n}\hat{\mathbf{M}}_{n-1|n-1}\mathbf{A}^{H}_{n}+\mathbf{C}_{\mathbf{r}_{n}}$ \\  \vspace{5pt}
$\mathbf{G}_{n}=\hat{\mathbf{M}}_{n|n-1}\mathbf{H}^{H}_{n}\big(\mathbf{H}_{n}\hat{\mathbf{M}}_{n|n-1}\mathbf{H}^{H}_{n}+\mathbf{C}_{\mathbf{s}^{a}_{n}}\big)^{-1}$ \\  \vspace{5pt}
$\hat{\mathbf{x}}_{n|n} = \hat{\mathbf{x}}_{n|n-1} + \mathbf{G}_{n}\big(\mathbf{y}_{n}-g(\hat{\mathbf{x}}_{n|n-1})\big)$ \\  \vspace{5pt}
$\hat{\mathbf{M}}_{n|n}=\left(\mathbf{I}-\mathbf{G}_{n}\mathbf{H}_{n}\right)\hat{\mathbf{M}}_{n|n-1}$ 
\label{Al:QEKF}
\end{algorithm}

\section{Quaternion Frequency Estimator}

The three-phase voltages in (\ref{eq:thee-phase voltages}) are now combined together to generate the pure quaternion signal
\[
\begin{aligned}
q_{n} = &  iv_{a,n}+jv_{b,n}+kv_{c,n}
\\
= &  iV_{a,n}\text{sin}\left(2\pi f \Delta T n + \phi_{a,n}\right)
\\
& +  jV_{b,n}\text{sin}\Big(2\pi f \Delta T n +  \phi_{b,n} +  \frac{2 \pi}{3}\Big)
\\
& +  kV_{c,n}\text{sin}\Big(2\pi f \Delta T n +  \phi_{c,n} +  \frac{4 \pi}{3}\Big).
\end{aligned}
\]
Through applying simple mathematical manipulations, the expression above yields
\begin{equation}
q_{n} = \Gamma_{I,n}\text{cos}(2 \pi \Delta T n) + \Gamma_{Q,n}\text{sin}(2 \pi \Delta T n)
\label{eq:q-IQ}
\end{equation}
where $\Gamma_{I,n}$ and $\Gamma_{Q,n}$ are given by
\[
\begin{aligned}
\Gamma_{I,n}=& iV_{a,n}\text{sin}(\phi_{a,n}) + jV_{b,n}\text{sin}(\phi_{b,n}+\frac{2 \pi}{3}) \\
&+ kV_{c,n}\text{sin}(\phi_{c,n}+\frac{4 \pi}{3})
\end{aligned}
\]
\[
\begin{aligned}
\Gamma_{Q,n}=& iV_{a,n}\text{cos}(\phi_{a,n}) + jV_{b,n}\text{cos}(\phi_{b,n}+\frac{4 \pi}{3}) \\
&+ kV_{c,n}\text{cos}(\phi_{c,n}+\frac{4 \pi}{3}).
\end{aligned}
\]
Replacing $\text{sin}(\cdot)$ and $\text{cos}(\cdot)$ functions with their polar representations, gives
\begin{equation}
\begin{aligned}
q_{n} = & \frac{\Gamma_{I,n}}{2}\big(e^{(\gamma 2 \pi f \Delta T n )} + e^{-( \gamma 2 \pi f \Delta T n )}\big)
\\
&+ \frac{\Gamma_{Q,n}}{2\gamma}\big( e^{(\gamma 2 \pi f \Delta T n )} - e^{- (\gamma 2 \pi  f \Delta T n )}\big) 
\end{aligned}
\label{eq:midway}
\end{equation}
where $\gamma= \Gamma_{I,n}\times \Gamma_{Q,n} / | \Gamma_{I,n}\times \Gamma_{Q,n} |$.

\begin{rem}
Although any pure quaternion number with unit amplitude can be chosen for $\gamma$; however, our choice  will simplify the obtained state space model and will give physical meaning to the state vector components.
\label{Rem:gamma}
\end{rem}

\begin{rem}
From (\ref{eq:q-IQ}), notice that $\Gamma_{I,n}$ and $\Gamma_{Q,n}$ lie in the same plane as $q_{n}$; hence, $\gamma$ is orthonormal to plane containing $q_{n}$.
\label{Rem:plane}
\end{rem}

The expression in (\ref{eq:midway}) is rearranged to give
\begin{equation}
\begin{aligned}
q_{n}= & \underbrace{\Big( \frac{\Gamma_{I,n}}{2} + \frac{\Gamma_{Q,n}}{2\gamma} \Big) e^{(\gamma 2 \pi f \Delta T n )}}_{q^{+}_{n}}
\\
& + \underbrace{\Big( \frac{\Gamma_{I,n}}{2} - \frac{\Gamma_{Q,n}}{2\gamma} \Big)  e^{-(\gamma 2 \pi f \Delta T n )}}_{q^{-}_{n}}
\end{aligned}
\label{eq:positive-negative} 
\end{equation}
where $q_{n}$ has been divided into  the two counter rotating signals $q^{+}_{n}$ and $q^{-}_{n}$, which can be expressed by the quaternion linear regressions
\begin{equation}
q^{+}_{n}= q^{+}_{n-1}e^{\gamma 2 \pi f \Delta T} \hspace{0.5em} \text{and} \hspace{0.5em} q^{-}_{n}=q^{-}_{n-1} e^{-\gamma 2 \pi f \Delta T}.
\label{eq:quaternion regressive}
\end{equation}

Taking into account the quaternion linear regressions in (\ref{eq:quaternion regressive}), where the phase incrementing element  of $q^{+}_{n}$ is the quaternion conjugate of the phase incrementing element of $q^{-}_{n}$,  a state space model for $q_{n}$ is proposed in Algorithm \ref{Al:Q-SS}, where $\varphi_{n} = e^{\gamma 2 \pi f \Delta T }$, $\mathbf{r}_{n}$ is the state evolution noise, and $s_{n}$ is the observation noise. Note that Algorithm~\ref{Al:Q-SS} can be implemented by employing the QEKF presented in~\cite{QKF}.

\begin{algorithm}[!h]
\caption{Quaternion-valued state space model (Q-SS)}
State evolution equation: $\begin{bmatrix}\varphi_{n}\\ q^{+}_{n}\\ q^{-}_{n}\end{bmatrix}=\begin{bmatrix}\varphi_{n-1}\\  q^{+}_{n-1}\varphi_{n-1}\\  q^{-}_{n-1}\varphi^{*}_{n-1}\end{bmatrix}+\mathbf{r}_{n}$ \vspace{0.5em}
\\
Observation equation: $q_{n}=\begin{bmatrix}0&1&1\end{bmatrix}\begin{bmatrix}\varphi_{n}\\ q^{+}_{n}\\ q^{-}_{n}\end{bmatrix}+ s_{n}$ \vspace{0.5em}
\\  \vspace{5pt}
Estimate of frequency: $\hat{f}_{n}=\frac{1}{2 \pi \Delta T} \Im \left( \text{ln}\left(\varphi_{n} \right) \right)$\\ 
\vspace{-8pt}
\label{Al:Q-SS}
\end{algorithm}

\begin{figure}[!h]
\centering
\includegraphics[width=9cm,trim = 0cm 1cm 0cm 0cm]{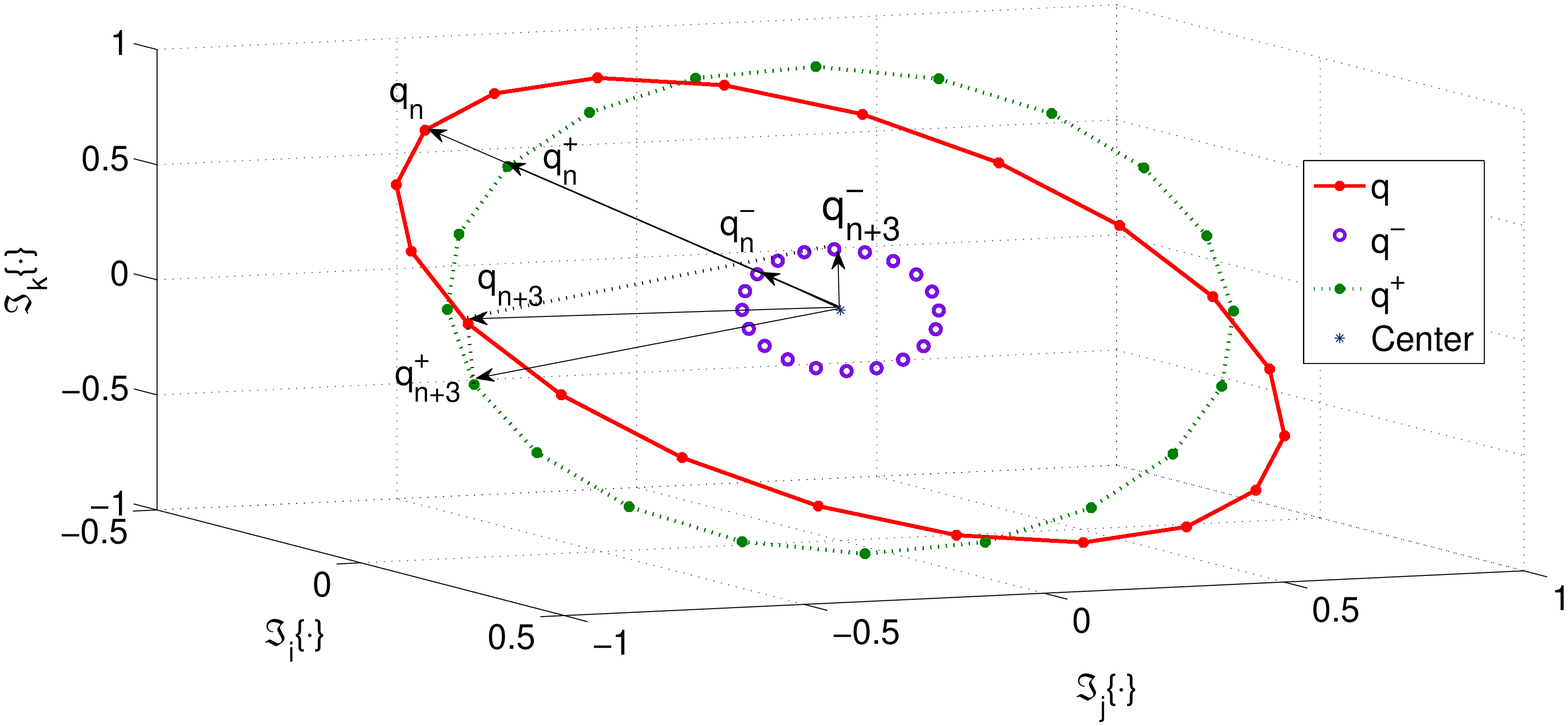}
\caption{System voltage, $q_{n}$, positive sequenced element, $q^{+}_{n}$, and negative sequence element, $q^{-}_{n}$, of an unbalanced three-phase system suffering from a $80$\% drop in the amplitude of $v_{a,n}$ and $20$ degree shifts in the phases of $v_{b,n}$ and $v_{c,n}$.}
\label{fig:positive and negative}
\end{figure}

\section{Phasor Estimation}

In order to be able to detect, characterize, and take appropriate action to mitigate voltage sags and faults in the power grid it is essential to monitor the relative voltage phasor information of each phase. Thus, we next estimate the voltage phasors of the system relative to phase $a$. Without loss of generality we assume $\phi_{a,n}=0$, this yields
\[
\begin{aligned}
\Gamma_{Q,n}&= iV_{a,n} + jV_{b,n}\text{cos}(\phi_{b,n}+\frac{4 \pi}{3})+ kV_{c,n}\text{cos}(\phi_{c,n}+\frac{4 \pi}{3})
\\
\Gamma_{I,n}&= jV_{b,n}\text{sin}(\phi_{b,n}+\frac{2 \pi}{3}) + kV_{c,n}\text{sin}(\phi_{c,n}+\frac{4 \pi}{3})
\end{aligned}
\]
where it becomes clear that $\Gamma_{Q,n}$ and $\Gamma_{I,n}$ consist of the real and imaginary components of the voltage phasors; thus, the problem of estimating the systems voltage phasors is reduce to estimating $\Gamma_{Q,n}$ and $\Gamma_{I,n}$. Notice that multiplying $q_{n}$ by $v_{a,n}$ gives
\[
\begin{aligned}
q_{n}v_{a,n}=& \frac{1}{2}\Gamma_{Q,n}V_{a,n} - \frac{1}{2}\Gamma_{Q,n}V_{a,n}\text{cos}(4\pi f \Delta T n) \\ & + \frac{1}{2}\Gamma_{Q,n}V_{a,n}\text{sin}(4 \pi f \Delta T n).
\end{aligned}
\]
Thus, $\Gamma_{Q,n}V_{a,n}$ can be estimated by passing $q_{n}v_{a,n}$ through a low pass filter (L.P.F.) which yields 
\[
\text{$2q_{n}v_{a,n}$ $\rightarrow$\begin{picture}(0,0)\put(0,-4){\framebox(30,15){L.P.F.}}\end{picture} \hspace{0.94cm}$\rightarrow$ $h_{n}=\Gamma_{Q,n}V_{a,n}$}
\]
It was assumed that $\phi_{a,n}=0$; therefore, it follow that
\begin{equation}
\Gamma_{Q,n} = \frac{h_{n}}{\sqrt{\Im_{i}(h_{n})}}\cdot
\end{equation} 

Taking into account the expression in (\ref{eq:positive-negative}) and applying tedious mathematical manipulations it can be shown that
\[
|q^{+}_{n}|^2 - |q^{-}_{n}|^2=|\Gamma_{I,n}||\Gamma_{Q,n}|\Re(\gamma_{I,n}\gamma_{Q,n}\gamma)
\]
where $\gamma_{I,n}=\Gamma_{I,n}/|\Gamma_{I,n}|$ and $\gamma_{Q,n}=\Gamma_{Q,n}/|\Gamma_{Q,n}|$. Considering from Remark~\ref{Rem:plane} that $\gamma \perp \{\gamma_{I,n}\text{ , }\gamma_{Q,n}\}$ yields
\[
\gamma_{I,n}\gamma_{Q,n}\gamma = \underbrace{(-\gamma_{I,n}\cdot \gamma_{Q,n})\gamma}_{\text{pure quaternion}} + \underbrace{(\gamma_{I,n}\times \gamma_{Q,n})\gamma}_{-\text{sin}(\phi_{I,Q})}
\]
where $\phi_{I,Q}$ is the angle between $\Gamma_{I,n}$ and $\Gamma_{Q,n}$; therefore, 
\begin{equation}
|q^{+}_{n}|^2 - |q^{-}_{n}|^2=-|\Gamma_{I,n}||\Gamma_{Q,n}|\text{sin}(\phi_{I,Q}).
\label{eq:state}
\end{equation}
Furthermore, using analytical geometrical it can be show that
\begin{equation}
\begin{aligned}
\Gamma_{Q,n}\times \Gamma_{I,n}=&-|\Gamma_{I,n}||\Gamma_{Q,n}|\text{sin}(\phi_{I,Q})\gamma 
\\
=& iV_{b,n}V_{c,n}\text{sin}(\frac{4 \pi}{3} + \phi_{c,n} - \frac{2 \pi}{3} - \phi_{b,n} ) 
\\
&- jV_{a,n}V_{c,n}\text{sin}(\frac{4\pi}{3}+\phi_{c,n}) 
\\
&+ kV_{a,n}V_{b,n}\text{sin}(\frac{2\pi}{3}-\phi_{b,n})
\end{aligned}
\label{eq:analytic}
\end{equation}
and therefore replacing (\ref{eq:state}) into (\ref{eq:analytic}) yields
\begin{equation}
\begin{aligned}
\Gamma_{I,n} =& -j \Im_{k}\bigg(\big(|q^{+}_{n}|^2-|q^{-}_{n}|^2\big)\gamma\bigg)\bigg(\frac{1}{\sqrt{\Im_{i}(h_{n})}}\bigg) 
\\ & + k \Im_{j}\bigg(\big(|q^{+}_{n}|^2-|q^{-}_{n}|^2)\gamma\bigg)\bigg(\frac{1}{\sqrt{\Im_{i}(h_{n})}}\bigg)
\end{aligned}
\end{equation}
where $\gamma$ can be replaced with its estimate $\Im(\rho_{n})/|\Im(\rho_{n})|$.

\section{Simulations} % Figure ratios 68.9990 and 1122.44 and size 20 Q-SS red WL-SS blue L-SS green

In this section, the performance of the developed quaternion frequency estimator is validated and compared to that of its complex-valued counterparts. In all experiments, the sampling frequency was $f_{s}=1$~KHz and the voltage measurements were considered to be corrupted by white Gaussian noise with signal to noise ratio (SNR) of $40$~dB. 

\begin{figure}[!h]
\centering
\includegraphics[width=1\linewidth, trim = 0cm 0cm 0cm 0.8cm]{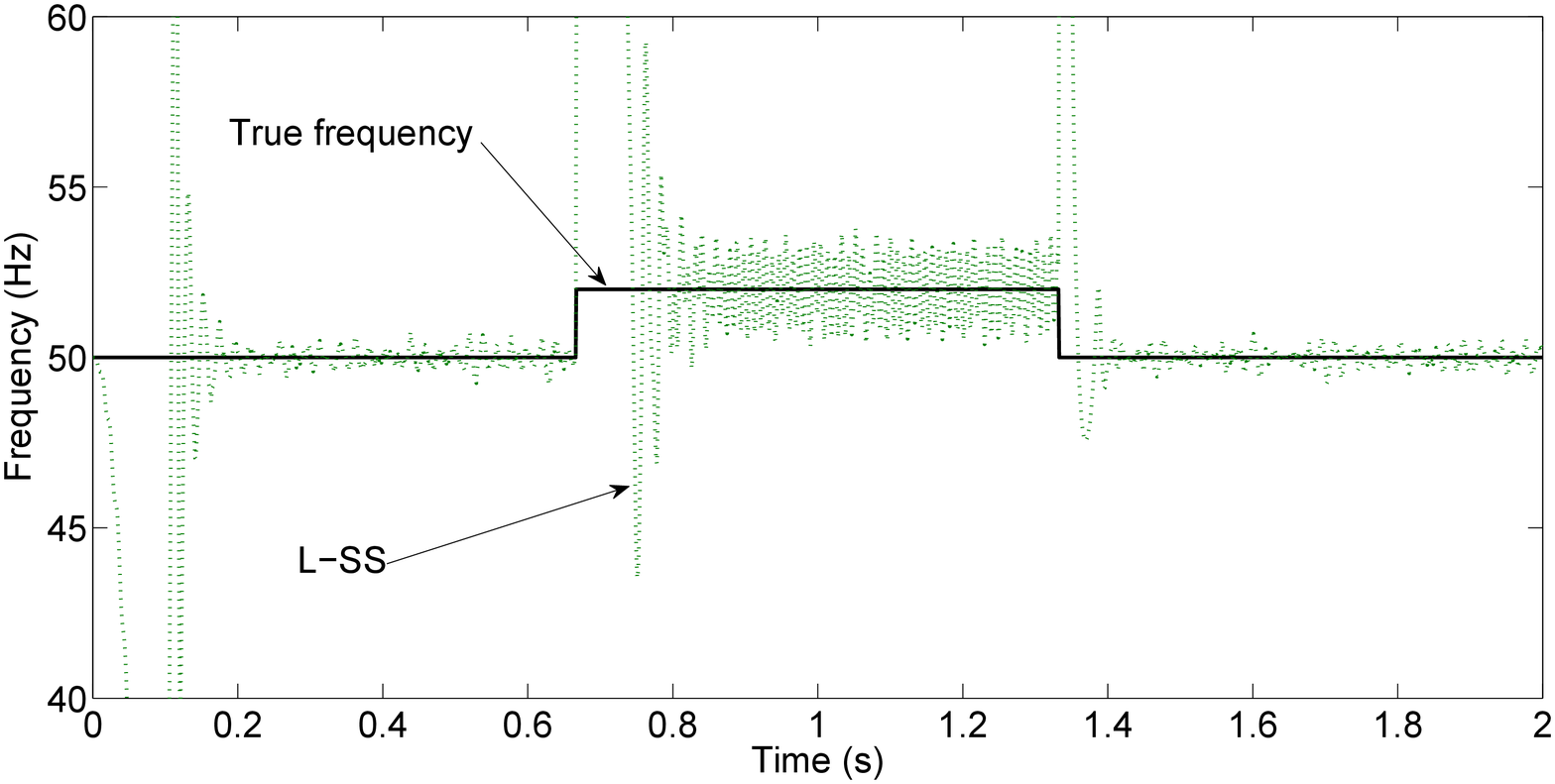}
\\
\includegraphics[width=1\linewidth, trim = 0cm 1cm 0cm 0cm]{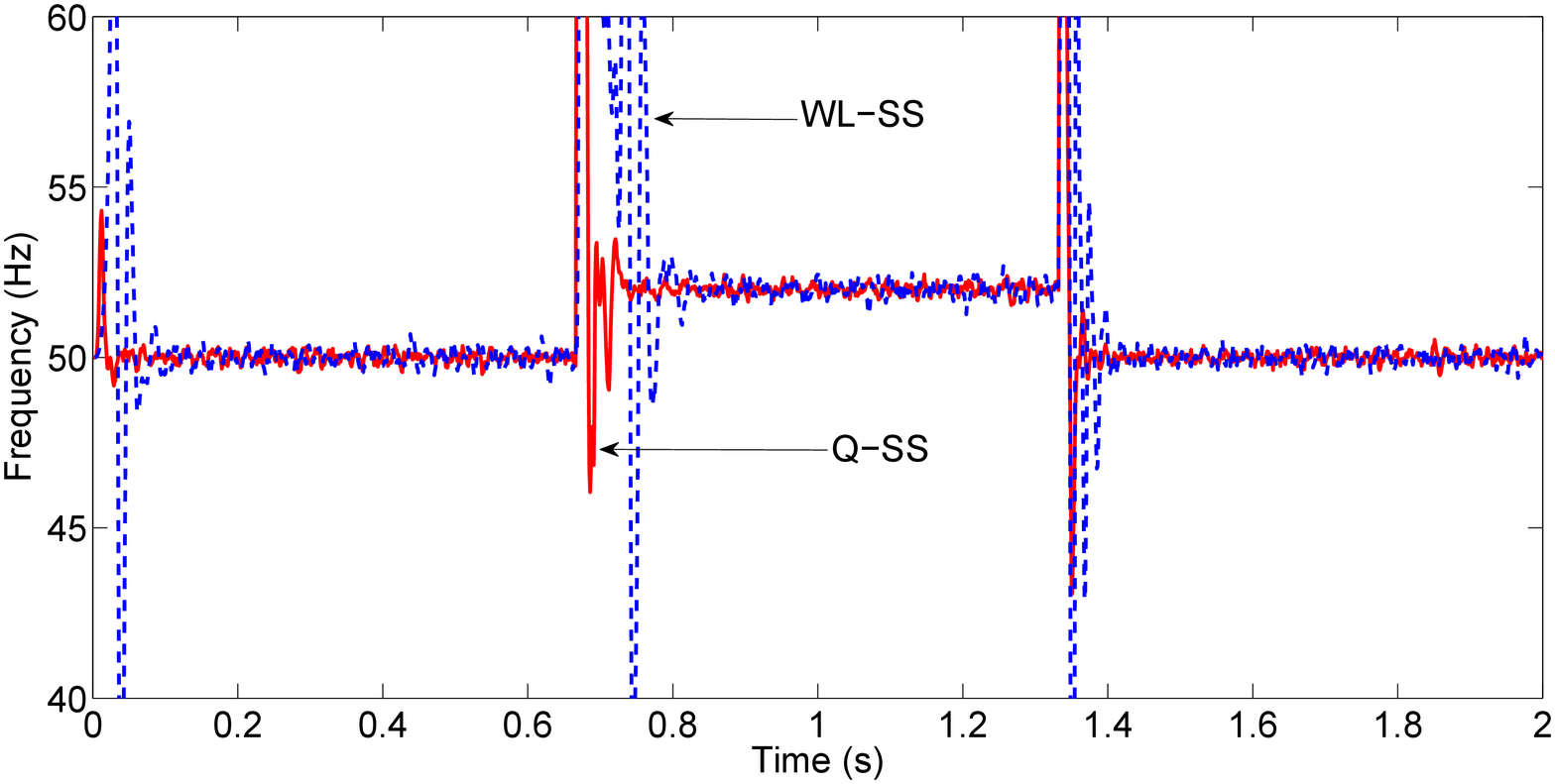}
\caption{Frequency estimation for a three-phase power system experiencing a short voltage sag and a $2$~Hz jump in frequency from $0.667$ to $1.334$ seconds. The performance of the L-SS algorithm is shown in the top graph and the performance of the WL-SS and Q-SS algorithms are compared in the bottom graph.}
\label{fig:freq-three}
\end{figure}

\begin{figure}[!h]
\includegraphics[width=1\linewidth, trim = 0cm 0cm 0cm 0.8cm]{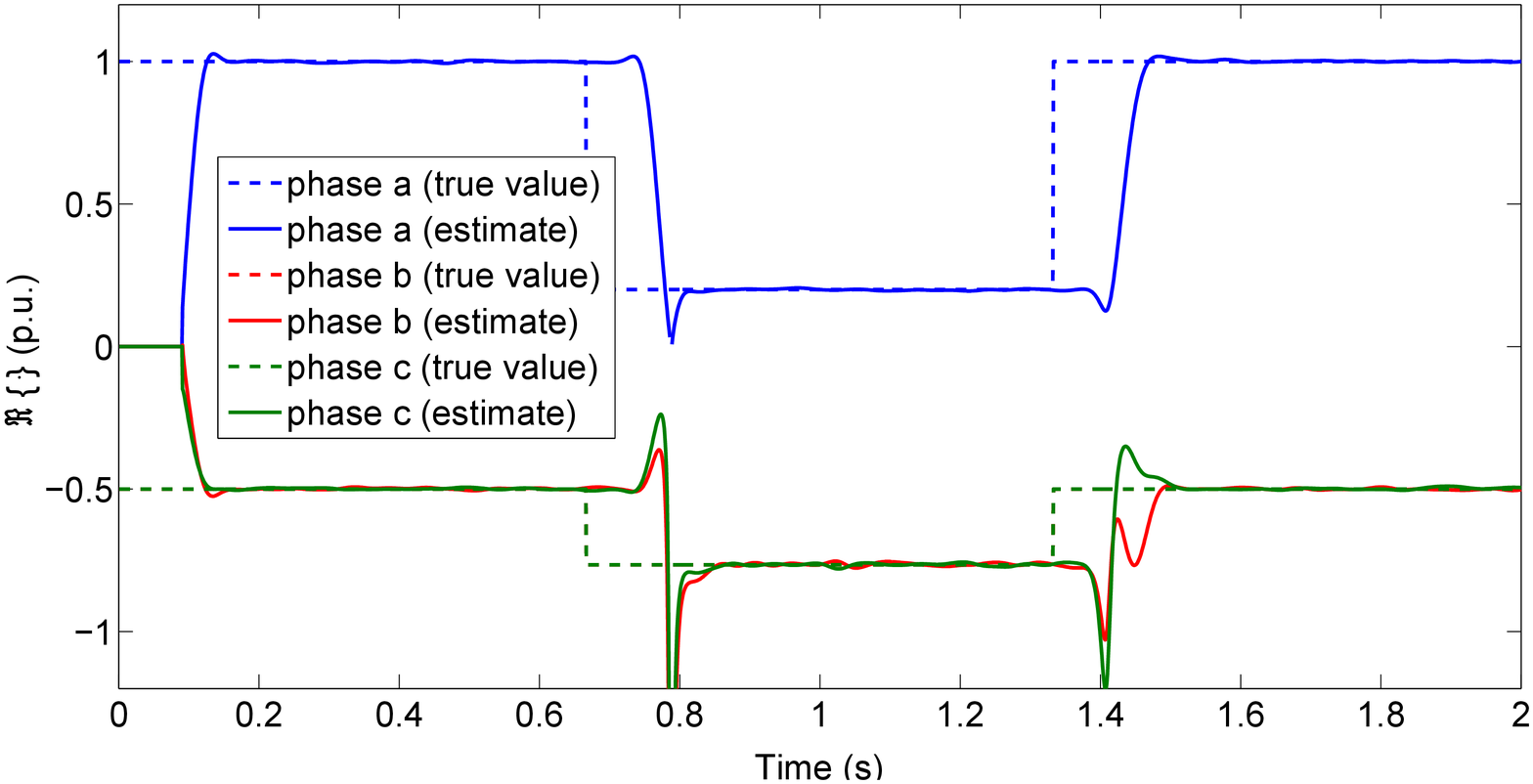}
\\
\includegraphics[width=1\linewidth, trim = 0cm 1cm 0cm 0cm]{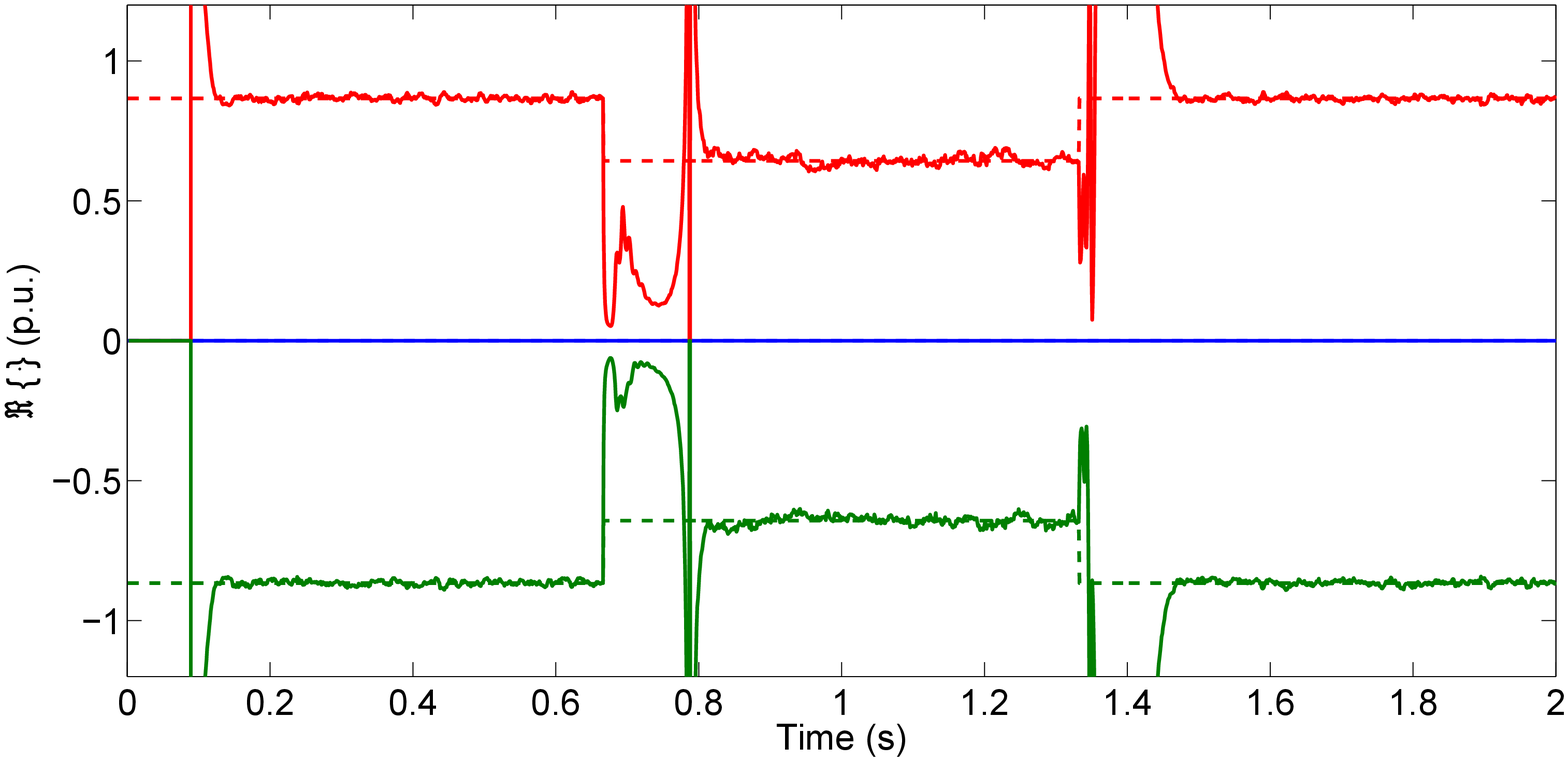}
\caption{Voltage phasor estimation for a three-phase system experiencing a short voltage sag from $0.667$ to $1.334$ seconds. The real components of the voltage phasors are shown in the top graph and the imaginary components of the voltage phasors are shown in the bottom graph.}  
\label{fig:phase-three}
\end{figure}

In the first experiment, the three-phase system was considered to be initially operating at its nominal frequency of $50$~Hz and in a balanced fashion, then the system suffers a voltage sag characterized by an $80$\% drop in the amplitude of $v_{a,n}$ and $20$ degree shifts in the phases of $v_{b,n}$ and $v_{c,n}$ (see Figure~\ref{fig:positive and negative}); furthermore, the frequency of the system experienced a step jump of $2$~Hz. The voltage sag lasted for a short duration and the system returned to balanced operating conditions and its nominal frequency once more. The estimates of the system frequency obtained through employing the L-SS, WL-SS, and Q-SS algorithms are shown in Figure~\ref{fig:freq-three}. Observe that the L-SS algorithm could only accurately estimate the system frequency during balanced operating conditions whereas the WL-SS and Q-SS algorithms tracked the system frequency during both balanced and unbalanced operating conditions. Moreover, the estimates of the system voltage phasors are shown in Figure~\ref{fig:phase-three}.

\begin{figure}[!h]
\includegraphics[width=1\linewidth, trim = 0cm 1cm 0cm 0cm]{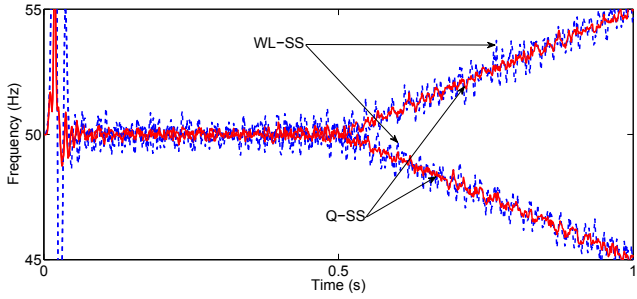}
\caption{Frequency estimation for an unbalanced three-phase system with changing frequency. Estimates obtained through implementing the Q-SS algorithm (in solid red line) and WL-SS (in dashed blue line) algorithms are compared.}
\label{fig:ramp}
\end{figure}

In the second experiment, we consider a three-phase system operating under unbalanced conditions caused by an $80$\% drop in the amplitude of $v_{a,n}$ and a $20$ degree shift in the phases of $v_{b,n}$ and $v_{c,n}$ (as shown in Figure~\ref{fig:positive and negative}), which experiences a ricing (\textit{cf}. falling) frequency due to mismatch between power generation and consumption $0.5$ seconds after the simulation starts. Figure~\ref{fig:ramp} shows the estimates of the system frequency. Observe that both when the system frequency was constant and when the system frequency was changing, the WL-SS and Q-SS algorithms accurately tracked the system frequency, with the Q-SS algorithm outperforming the WL-SS algorithm in terms of steady-state variance.  

\begin{figure}[!h]
\includegraphics[width=1\linewidth, trim = 0cm 0cm 0cm 1.5cm]{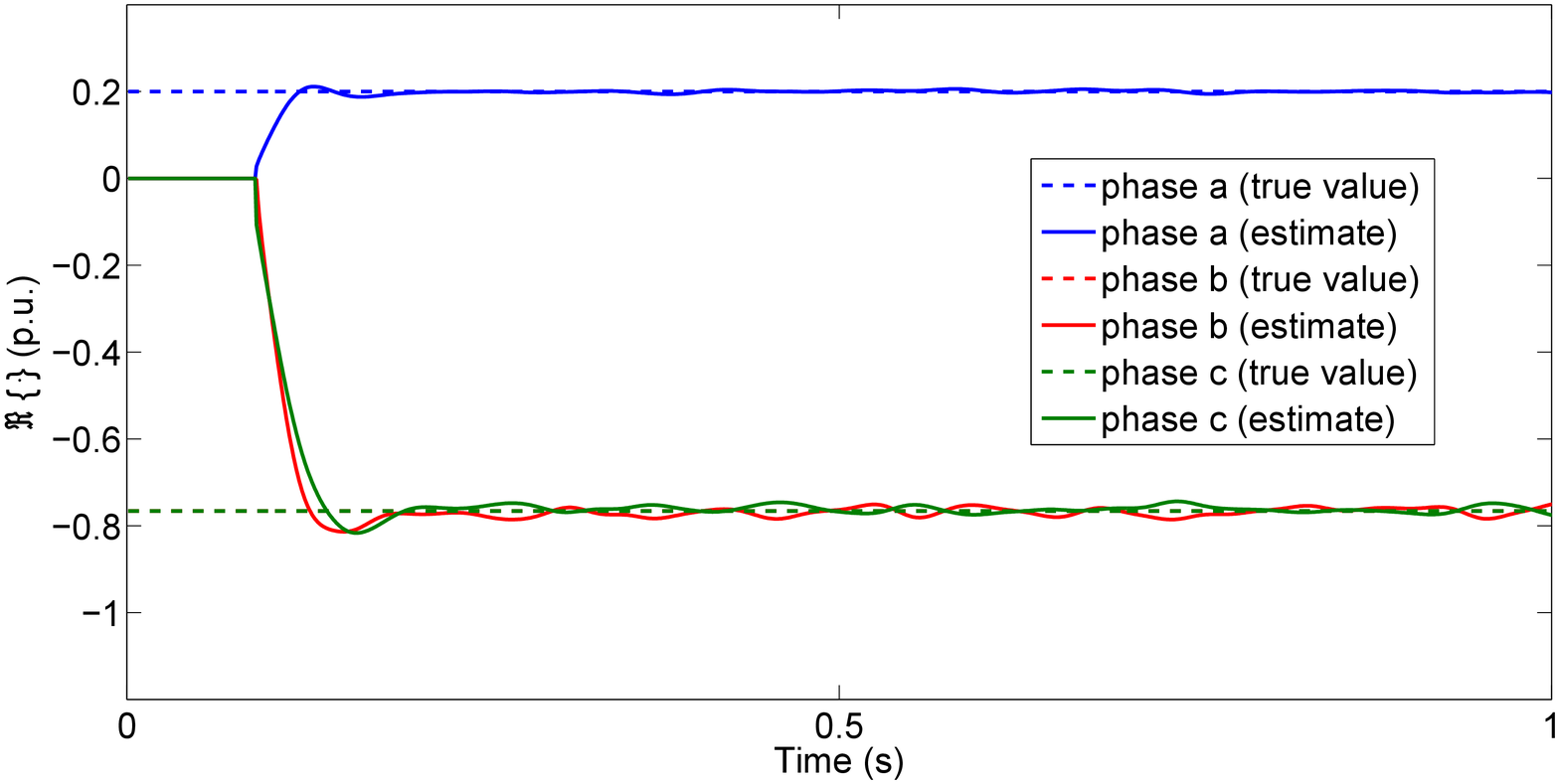}
\\
\includegraphics[width=1\linewidth, trim = 0cm 1cm 0cm 0cm]{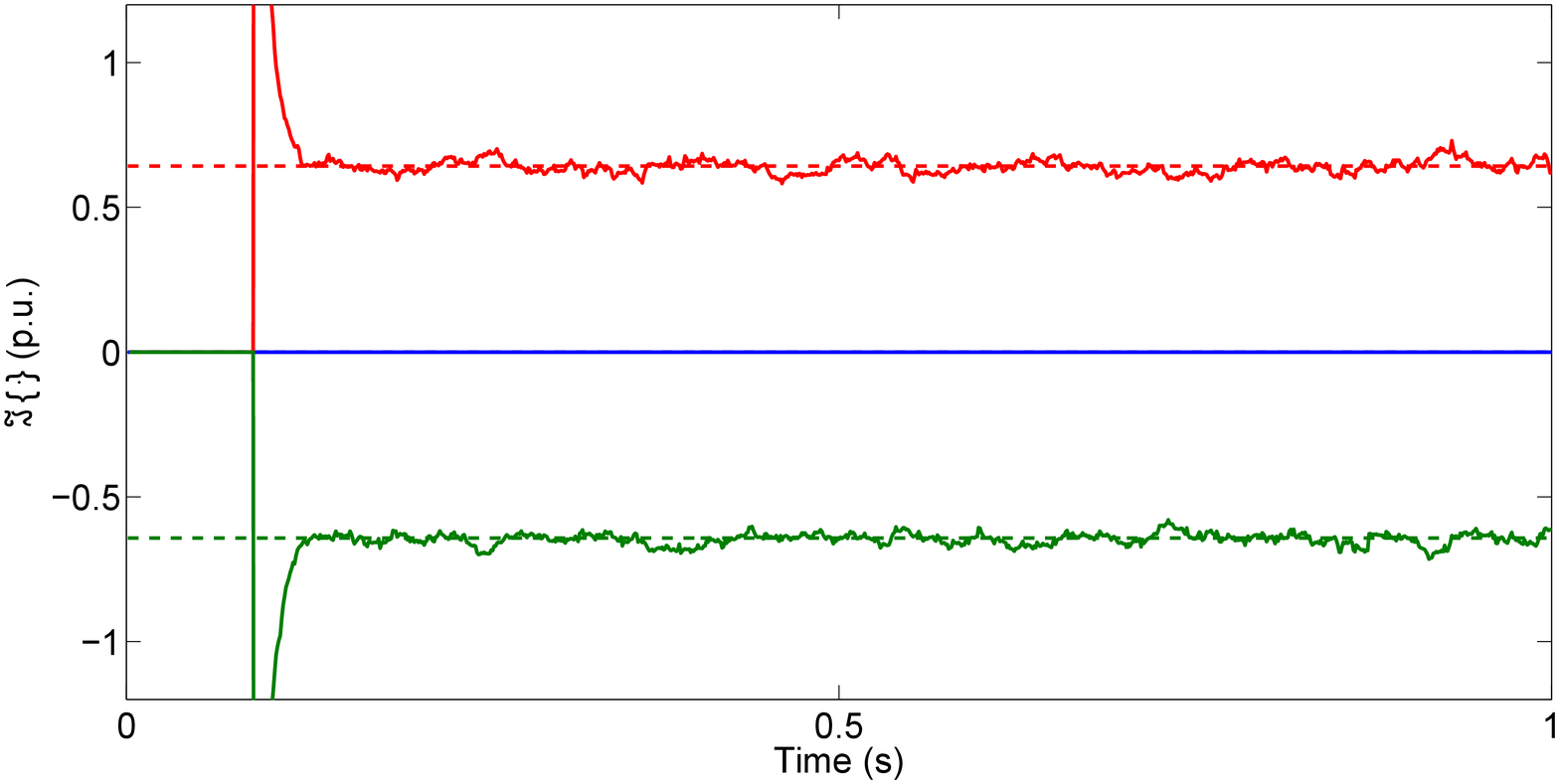}
\caption{Voltage phasor estimation for an unbalanced three-phase system with changing frequency. The real components of the voltage phasors are shown in the top graph and the imaginary components of the voltage phasors are shown in the bottom graph.}
\label{fig:ramp-phase}
\end{figure}

\begin{figure}[!h]
\includegraphics[width=1\linewidth, trim = 0cm 0.5cm 0cm 0cm]{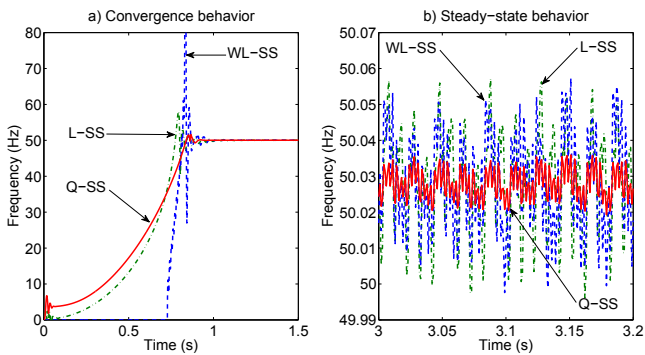}
\caption{Convergence and steady-state behavior of the Q-SS algorithm when dealing with real-world data where the system frequency was $50.028$~Hz: a)~convergence behavior, b)~steady-state behavior.}
\label{fig:Conv-Steady}
\end{figure}

In the third simulation, the convergence and steady-state behavior of the Q-SS algorithm was compared to those of the WL-SS and L-SS algorithms using real-world data. The obtained results are shown in Figure~\ref{fig:Conv-Steady}. Observe that although all the algorithms converged at the same time instance, the steady-state behavior of the quaternion-valued Q-SS algorithm is significantly better than those of the complex-valued WL-SS and L-SS algorithms.

\begin{figure}[!h]
\includegraphics[width = 1\linewidth, trim = 0cm 0cm 0cm 0cm]{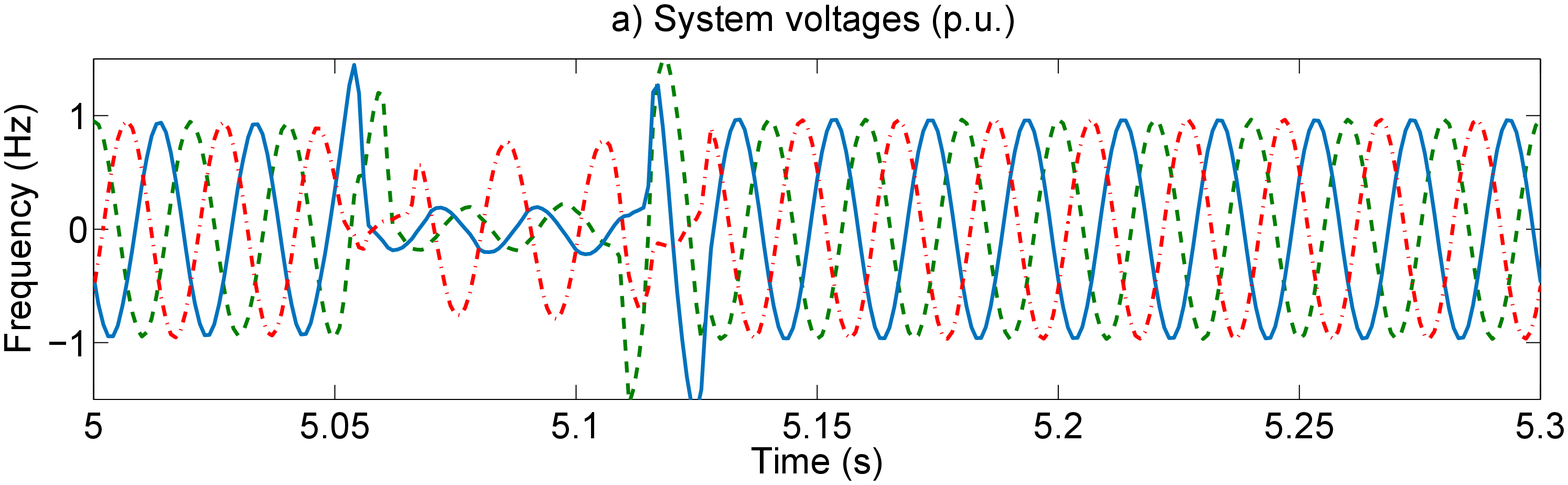}
\\
\includegraphics[width = 1\linewidth, trim = 0cm 0.5cm 0cm 0cm]{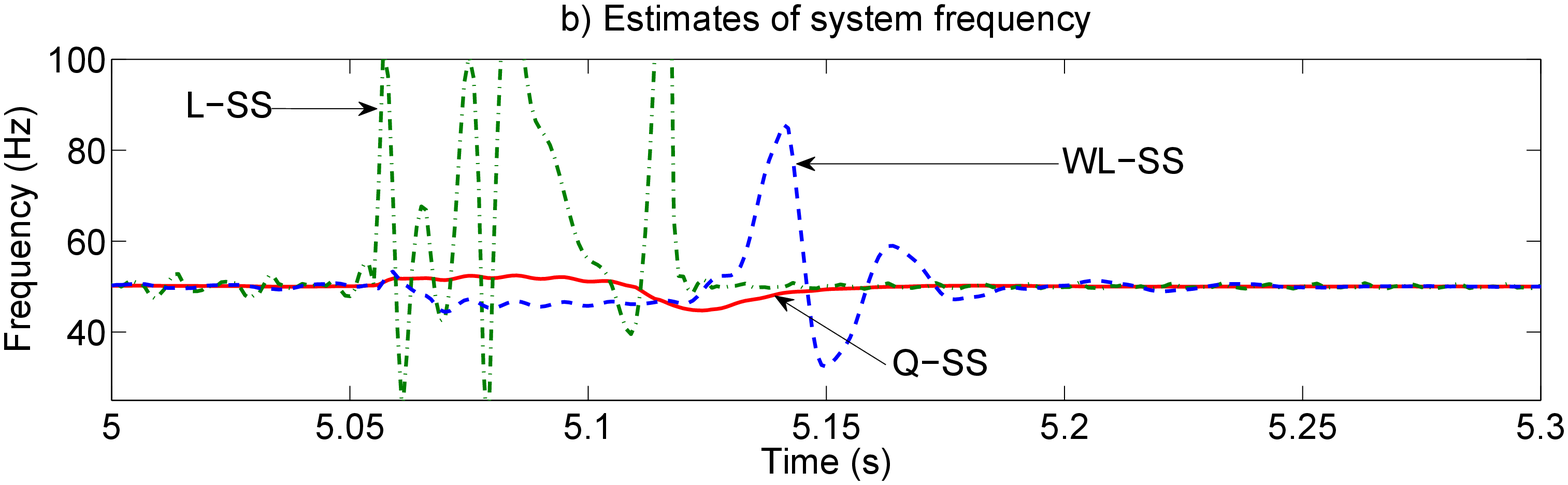}
\caption{Frequency estimation using real-world data recording during a voltage sag: a) system voltages, b) estimates of the system frequency obtained through implementing the Q-SS, WL-SS and L-SS algorithms.}
\label{fig:real-sag}
\end{figure}

In the forth simulation, the performance of the developed Q-SS algorithm was assessed during a voltage sag using real-world data, where  the system suffered a voltage sag $5.4$ seconds after recording started, which lasted for $80$ milliseconds. The recorded data and the performance of the Q-SS, WL-SS, and L-SS algorithms are shown in Figure~\ref{fig:real-sag}. Notice that the L-SS algorithm lost track of the system frequency during the voltage sag and that the WL-SS algorithm suffered from a large momentary error once the voltage sag was over; however, the Q-SS algorithm showed outstanding performance both during the voltage sag and when the voltage sag was over.       

\begin{figure}[!h]
\includegraphics[width = 1\linewidth, trim = 0cm 1cm 0cm 0cm]{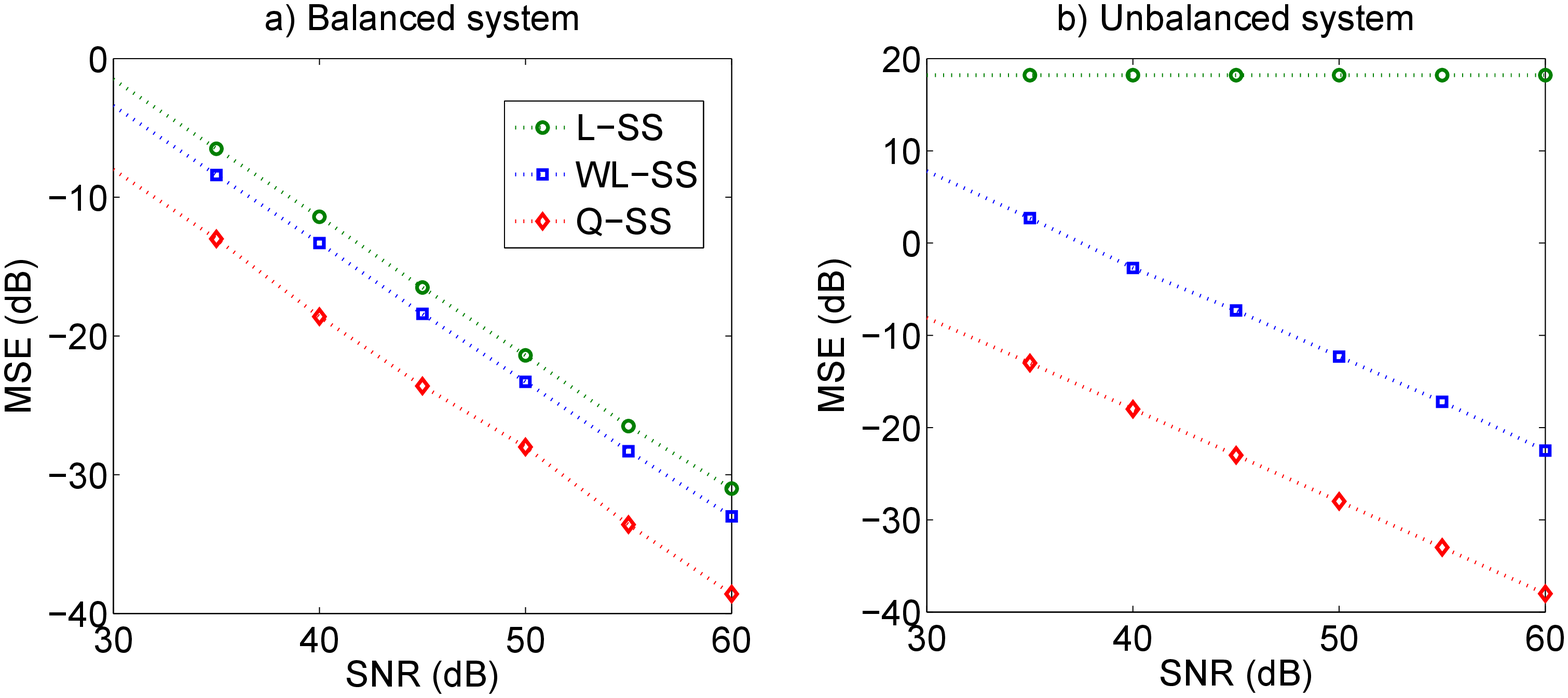}
\caption{MSE performance of the Q-SS and DQ-SS algorithms: a) balanced three-phase system, b) unbalanced three-phase system suffering from  an $80$\% drop in the amplitude of $v_{a,n}$ and $20$ degree the shifts in the phases of $v_{b,n}$ and $v_{c,n}$.}
\label{fig:MSEperformance}
\end{figure}

Finally, the MSE performance of the developed algorithm for both a balanced and an unbalanced power system suffering from an $80$\% drop in the amplitude of $v_{a,n}$ and $20$ degree the shifts in the phases of $v_{b,n}$ and $v_{c,n}$  are shown in Figure~\ref{fig:MSEperformance}. Notice that the newly developed Q-SS algorithm outperforms the WL-SS and L-SS algorithms and in contrast to the WL-SS ad L-SS algorithms, imbalanced operating conditions does not effect the MSE performance of the Q-SS algorithm, a desirable characteristic for frequency estimators in three-phase systems; furthermore, employing the developed frequency estimator in its distributed form, DQ-SS, further reduced the MSE by $4$dB.

\section{Conclusion}

A novel algorithm based on the extended quaternion Kalman filter (QEKF) and the $\mathbb{HR}$-calculus has been developed for estimating the fundamental frequency of three-phase power systems. The proposed algorithm exploits the multidimensional nature of quaternions to make possible the full characterization of three-phase power systems in the three-dimensional domain, where they naturally reside and eliminates the need for using the Clarke transform. The performance of the proposed algorithm has been assessed in a number of scenarios using both synthetic and real-world data, where it has shown outstanding performance and outperformed its complex-valued counterparts. 

\balance

\end{document}